\documentclass[journal]{IEEEtran}
\usepackage[english = american]{csquotes} 
\usepackage{url,multirow,comment,hyperref,amsmath,graphicx,amsmath,subfigure,listings,tabularx,psfrag,wrapfig,cite,verbatim,array, algorithm, algorithmicx, algpseudocode}
\usepackage[
locale = DE 
]{siunitx} 
\usepackage{caption} 
\usepackage{xcolor}
\usepackage{afterpage}
\usepackage{float} 
\usepackage{caption} 
\begin{document}

\title{Design and Verification of a Synchronus First In First Out (FIFO) }

\author{
Yatheeswar~Penta~and Riadul~Islam,~\IEEEmembership{Senior Member,~IEEE}


\thanks{Y. Penta and R Islam are with the Department 
of Computer Science and Electrical Engineering, University of Maryland, Baltimore County, 
MD 21250, USA e-mail: {riaduli@umbc.edu}.}
\thanks{}
}

\markboth{IEEE Transactions on Computer-Aided XXXXXX}%
{Shell \MakeLowercase{\textit{et al.}}: ??????}

\newcommand{\fixme}[1]{{\Large FIXME:} {\bf #1}}

\maketitle

\begin{abstract}

This project focuses on designing and verifying a synchronous FIFO First In First Out (FIFO) memory, a critical component in digital systems for temporary data storage and seamless data transfer. The FIFO operates under a single clock domain, ensuring synchronized read and write operations, making it suitable for systems requiring high-speed, reliable data buffering. This design includes FIFO's key features, such as read and write operations, full and empty flag generation, and pointer management for memory control. The FIFO was implemented using Verilog to define the Register Transfer Level (RTL) design, ensuring functionality and timing requirements were met. 
For verification, three approaches were employed: (1) UVM-based Verification: A Universal Verification Methodology (UVM) testbench was developed to test the FIFO design rigorously. The testbench includes components like interface, sequence item, driver, monitor, scoreboard, agent, and environment. Directed and random tests were performed to verify corner cases, such as simultaneous reads and writes, full and empty conditions, and overflow and underflow scenarios; (2) Traditional Verilog Testbench: A standalone Verilog testbench was also used to validate the functionality of the FIFO through directed test scenarios and waveform analysis; (3) FPGA implementation: Additionally, the design was implemented on an FPGA for real-time testing to verify its functionality and timing behavior in hardware. FPGA-based verification ensured the design performed as expected under practical conditions.
The results confirmed the correct operation of the FIFO, including accurate data transfer, flag behavior, and timing synchronization. The project successfully demonstrated the robustness and reliability of the synchronous FIFO design, highlighting its importance in modern digital systems for efficient data handling and buffering.

\end{abstract}


%
\IEEEpeerreviewmaketitle

\section{Introduction}
A FIFO (First-In, First-Out) is a type of data buffer or memory storage element used in digital systems to manage and organize data transfer~\cite{xie_enze:23}. As the name suggests, data that enters the FIFO first is the first to be read or processed, following a queue-based structure. FIFOs are widely used in digital design to handle data flow between components operating at different rates, ensuring smooth and efficient data management.

Key Features of a FIFO include (i) Queue Behavior: Data is written into the FIFO in sequence (write operation) and read in the same sequence (read operation); (ii) Full and Empty Flags: Full Flag indicates that the FIFO cannot accept more data.
Empty Flag indicates that there is no data available for reading~\cite{Ghosh:2023}; (iii) Read and Write Pointers: The FIFO internally maintains separate read and write pointers to track the memory locations for reading and writing operations~\cite{Ciletti:2010}; (iv) Overflow and Underflow: Overflow occurs if data is written into a full FIFO. Underflow occurs if a read operation is attempted on an empty FIFO.

FIFOs are generally classified into two types: Synchronous FIFOs, which operate using a single clock for both read and write operations. They are ideal for systems where both reading and writing occur in the same clock domain and asynchronous FIFOs, which use separate clocks for read and write operations. Suitable for scenarios where data transfer occurs between different clock domains~\cite{Islam_diff:2015, Kumar:2014, islam2018low, guthaus2017current, Islam_neg:2021}.

FIFOs are extensively used in modern digital systems for tasks like, Data Buffering: Temporarily storing data in communication systems and processors, Clock Domain Crossing (CDC)~\cite{Islam_hcdn:2019, Islam_dcmcs:2018, Lin:2015, islam2017current}: Managing data flow between components with different clock rates, Networking and Communication: Handling data packets in routers and switches, Image and Video Processing: Buffering continuous data streams for real-time processing, Embedded Systems~\cite{Croteau:2024, saini2024reconfigurable, Kiriakidis:2022, islam2022systems}: Managing data transfer between sensors, processors, and peripherals.
FIFOs play a vital role in improving system efficiency and ensuring smooth data flow in various real-time applications~\cite{Narkhede:2019}. Their ability to decouple components with different data rates and handle temporary storage makes them indispensable in digital systems like microcontrollers, FPGAs, and communication interfaces.

\subsection{VLSI Design and FIFO}
In digital systems, data transfer between components operating at different clock frequencies known as CDC is a common challenge in VLSI design. Ensuring reliable communication across these clock domains is critical to system performance and stability.~\cite{Nanda:2018} Traditionally, synchronizers such as flip-flops~\cite{Islam_sedu:2019, islam2011high, Islam_sdff:2011, Esmaeili:2012, Islam_isqed:2012} are used to align asynchronous signals with a local clock. However, these synchronizers can suffer from setup and hold time violations, which can lead to metastability~\cite{Bhanu:2021}. Metastability is an unpredictable state where the flip-flop fails to resolve to a definite logic level, causing timing errors and data corruption~\cite{5993629}. While synchronizers can mitigate some of these issues, they are not always the ideal solution for reliable data transfer, especially when large volumes of data are involved.

To address these challenges, synchronous FIFOs are widely employed in VLSI systems. Synchronous FIFOs act as buffers to manage data transfer in systems operating at the same clock frequency but different data rates~\cite{Rizvi:2015}. By decoupling the data producer (write operation) from the data consumer (read operation), FIFOs prevent issues like data overflow and underflow, ensuring smooth and efficient communication. Synchronous FIFOs inherently simplify timing closure as they operate under a single clock domain, eliminating the need for complex synchronizers to manage timing mismatches.

Furthermore, synchronous FIFOs provide robust flow control using status flags like Full and Empty, which signal when the FIFO is at capacity or when it has no valid data to read. This ensures that no invalid operations take place, maintaining data integrity~\cite{Sharma:2013}. Compared to basic synchronizers, synchronous FIFOs offer a more reliable and scalable solution for managing data flow, making them an indispensable component in high-speed digital systems, including processors, memory controllers, and communication interfaces. In addition, Design verification is an essential investment in microprocessor development, enabling companies to identify and address potential issues before they escalate into costly production and operational problems. By ensuring accuracy, reliability, and efficiency, verification reduces overall development costs, accelerates time-to-market, and enhances product competitiveness~\cite{islam2022feasibility, Islam_early:2022, challagundla2023design, islam2019predicting}. This research addresses this issue by designing FIFO and verifying using the FPGA and Universal Verification Methodology (UVM) approaches.


\section{Background}

\subsection{Synchronus FIFO}
A Synchronous FIFO is a memory buffer that allows sequential storage and retrieval of data, operating entirely within a single clock domain. Both read and write operations are synchronized to the same clock signal, ensuring consistent timing and eliminating the need for complex clock domain management~\cite{Wyland:1993, islam2021high}. The synchronous FIFO works on a queue-based principle, where data written first is the first to be read, maintaining an orderly flow of information.

In VLSI (Very Large Scale Integration) design, synchronous FIFOs play a critical role in ensuring reliable and efficient data handling. They serve as buffers that temporarily store data between functional units operating at different data rates within the same clock domain. By acting as intermediaries, FIFOs ensure smooth data transfer, preventing data loss or corruption caused by mismatched rates of data production and consumption. The single clock synchronization simplifies timing closure and eliminates issues such as metastability, which are common in multi-clock domain systems. This makes synchronous FIFOs easier to implement and analyze, especially in high-performance designs.~\cite{Maurya:2016}

Synchronous FIFOs are essential for managing high-speed data transfers with minimal latency. They allow systems to achieve higher throughput and ensure predictable timing behavior, which is critical for applications like processors, memory controllers, and communication systems. Built-in control flags, such as full and empty, simplify data flow management and provide reliable status signals, ensuring no invalid operations occur. This inherent flow control mechanism makes synchronous FIFOs highly robust and reliable in real-world designs.~\cite{Darda:2015}

Another key advantage of synchronous FIFOs is their modular and reusable nature. They are often used as building blocks in complex systems and can be integrated seamlessly into larger subsystems, such as System-on-Chip (SoC) designs. Their optimized implementation also makes them power- and area-efficient, which is crucial for resource-constrained and low-power applications. Additionally, synchronous FIFOs are widely used in pipelined architectures, where they help decouple stages to maintain efficient data flow and prevent bottlenecks.~\cite{Ahmed:2023}

The applications of synchronous FIFOs in VLSI design are extensive. They are used in microprocessors for buffering instructions and handling data transfer, in memory controllers to bridge the speed gap between processors and memory subsystems, and in communication systems to manage packet-based data transfers. Embedded systems also rely on synchronous FIFOs to facilitate data exchange between processors, peripherals, and sensors, ensuring real-time performance. Their simplicity and predictability make them easy to implement and verify using functional verification methods like UVM (Universal Verification Methodology).~\cite{Schirrmeister:2014} Furthermore, synchronous FIFOs can be quickly prototyped and validated on FPGAs, making them ideal for real-world testing and debugging.

\subsection{Working of Synchronus FIFO}
The operation of a synchronous FIFO is managed by read and write pointers that track the locations in the FIFO memory for data storage and retrieval. When data is written into the FIFO, the write pointer increments, pointing to the next available memory location. Similarly, when data is read from the FIFO, the read pointer increments, ensuring the next piece of data is retrieved in the correct sequence~\cite{Darbari:2016}.

To ensure smooth operation and avoid invalid data handling, the FIFO generates two key control flags (i.e., Full and Empty).
The Full flag indicates that the FIFO is at maximum capacity, and no more data can be written until space is freed. Conversely, the Empty flag signals that the FIFO contains no valid data, preventing invalid read operations. These flags are derived by comparing the positions of the read and write pointers. If the write pointer is one step ahead of the read pointer, the FIFO is declared full. If both pointers are equal, the FIFO is declared empty.

{\bf Write Operation:} During a write operation, the following steps are executed:
The write enable (wn) signal is asserted to indicate a valid write request.
The data is written into the memory location pointed to by the write pointer.
After the write operation, the write pointer increments to point to the next location.
If the write pointer wraps around and catches up to the read pointer, the Full flag is set to prevent further writes.~\cite{Jain:2022}

{\bf Read Operation:} The read enable (rn) signal is asserted to initiate a valid read request.
The data is read from the memory location pointed to by the read pointer.
After the read operation, the read pointer increments to point to the next location.
If the read pointer catches up with the write pointer, the Empty flag is set to prevent further reads.
The write and read pointers increment independently but are synchronized to the same clock, ensuring no timing mismatches. The pointers wrap around when they reach the end of the FIFO memory, creating a circular buffer~\cite{Liu:2021}.

The comparison of these pointers determines the status flags. 
Full: When the write pointer is one position ahead of the read pointer (circular wrap-around).
Empty: When the write pointer and read pointer are equal.
Overflow and Underflow:
Overflow occurs if a write operation is attempted when the FIFO is full. To prevent this, the full flag disables further writes.
Underflow occurs if a read operation is attempted when the FIFO is empty. The empty flag prevents invalid reads from occurring~\cite{Saxena:2018}.

The synchronous FIFO operates entirely on a single clock edge (rising or falling), ensuring that all read and write operations are synchronized. This eliminates timing issues, simplifies design, and allows easy integration into larger synchronous systems like processors and memory controllers.

\begin{figure}
    \centering
    \includegraphics[width=1\linewidth]{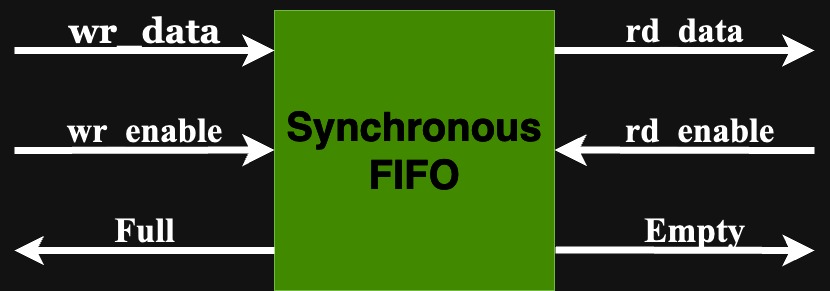}
    \caption{Basic diagram of Synchronus FIFO.}
    \label{fig:enter-label}
\end{figure}

\subsection{Design Specifications}
\textbf{Inputs:}

data-in: 8-bit input data signal

clock: 1-bit clock signal

reset: 1-bit reset signal (active high)

wn: 1-bit write enable signal

rn: 1-bit read enable signal

\textbf{Outputs:}

data-out: 8-bit output data signal

full: 1-bit flag indicating FIFO is full

empty: 1-bit flag indicating FIFO is empty

\textbf{Internal Signals}

wptr: 3-bit write pointer

rptr: 3-bit read pointer

memory: 8-entry array of 8-bit registers (FIFO storage)

\textbf{Behavior}

Reset: On a high reset signal, all entries in the FIFO memory are
cleared, wptr and rptr are reset to 0, and data-out is set to 0.

Write Operation: When wn is high and full is low, data-in is written into the FIFO at the position indicated by wptr. The wptr is then incremented.~\cite{Cheng:2019}

Read Operation: When rn is high and empty is low, data from the FIFO at the position indicated by rptr is output through data-out. The rptr is then incremented.

Full Condition: The FIFO is considered full when wptr[2:1] ==
rptr[2:1] and wptr[0] != rptr[0]. When full, no further write
operations are allowed.

Empty Condition: The FIFO is considered empty when wptr ==
rptr. When empty, no further read operations are allowed.

\textbf{Clocking }

Synchronous with the clock signal. Operations are triggered on the
positive edge of clock.

\textbf{Memory Depth}

8 entries (indexed by 3-bit pointers)

\textbf{Data Width}

8 bits per entry

\textbf{Use Cases}

Buffering: Temporarily stores data to handle timing differences
between producer and consumer modules.

Queue Management: Manages data flow in a first-in, first-out
manner.

\section{Proposed Methodology}
The FIFO was designed using Verilog to implement the RTL (Register Transfer Level) description, ensuring that both functionality and timing requirements were satisfied.  
For verification, we utilized three approaches: 

1. UVM-based Verification, 

2. Traditional Verilog Testbench, and 

3. FPGA Implementation. 

\subsection{Universal Verification Methodology (UVM)}

The Universal Verification Methodology (UVM) is a standardized verification methodology widely used in VLSI design for verifying complex digital systems.~\cite{Juan:2014} It provides a well-structured and reusable framework for creating verification environments using SystemVerilog. UVM leverages object-oriented programming (OOP) concepts such as classes, inheritance, and polymorphism, enabling scalable and efficient verification processes.~\cite{IEEE1800:2017}

\subsubsection{\textbf{Components of UVM}}

A UVM testbench consists of various verification components that are reusable, configurable, and modular. These components collectively form a verification environment that can efficiently stimulate, observe, and verify the Design Under Test (DUT). Each component serves a specific purpose, ensuring seamless interaction between test scenarios and the DUT. They are

\subsubsection{\textbf{Sequence and Sequencer in UVM}}

In UVM (Universal Verification Methodology), the Sequence and Sequencer work together to generate and deliver input stimulus to the Driver, which drives the signals to the DUT (Design Under Test). These components play a critical role in creating flexible and reusable testbenches.

\textbf{Sequence:} A Sequence is responsible for generating transactions or input stimulus for the DUT.
It defines the type and nature of the data to be sent, such as random stimulus, directed patterns, or constrained inputs. ~\cite{Mehta:2013}
Sequences are created by extending the uvm-sequence class.
A sequence can generate simple or complex scenarios, allowing verification engineers to simulate various corner cases and functional paths of the DUT~\cite{Kadam:2017}.

\textbf{Sequencer:} The Sequencer is responsible for scheduling and managing sequences.
It acts as a bridge between the Sequence and the Driver.
The sequencer passes the transactions generated by the sequence to the driver for execution.
The sequencer ensures a controlled flow of transactions and can handle multiple sequences at once.
It is created by extending the uvm-sequencer class~\cite{Juan:2014}.

\textbf{Working Principles:} The sequence generates a transaction (stimulus).
The sequencer schedules this sequence and delivers the transaction to the driver.
The driver translates the transaction into pin-level signals and drives it to the DUT.
Together, the sequence and sequencer provide a modular and flexible way to control stimulus generation in a UVM testbench. This separation of stimulus generation (sequence) and scheduling (sequencer) enables efficient verification and reuse across projects.

\subsubsection{\textbf{Driver in UVM}}
In Universal Verification Methodology (UVM), the Driver is an active verification component responsible for driving signals to the Design Under Test (DUT). Converts high-level transactions (generated by the sequencer) into low-level pin-level signal activity that the DUT can understand. The driver ensures that the DUT receives valid input stimulus in a controlled and organized manner.~\cite{Agarwal:2014}

\textbf{Role of the Driver:} The driver takes transactions (data packets or instructions) generated by the sequence and translates them into bit-level or signal-level operations.

Driving DUT applies the converted signal-level data to the DUT inputs through the virtual interface.
Then the driver communicates with the sequencer to receive transactions, processes them, and informs the sequencer when the transaction is completed.
The driver ensures that the timing and signal behavior align with the DUT's specifications, handling clocking requirements and setup times.

\textbf{Driver's Interaction in UVM Flow:} The Sequencer generates and sends transactions to the Driver. The Driver converts these transactions into low-level signal activity.
The Driver applies the signals to the DUT inputs using the virtual interface.

\subsubsection{\textbf{Monitor in UVM}}
In UVM, the Monitor is a passive verification component responsible for observing and capturing the signal activity on the interface of the DUT~\cite{Juan:2014}. Unlike the Driver, the Monitor does not drive any signals to the DUT; it solely monitors the behavior of the DUT inputs and outputs. The captured signal-level activity is converted into high-level transactions that can be passed to other components like the scoreboard for checking or the coverage collector for functional coverage analysis.

\textbf{Key Roles of the Monitor:} Signal Observation:
The monitor continuously observes input and output signals from the DUT through the virtual interface.
Transaction Creation:
The observed signal-level activity is converted into transaction-level data (e.g., packets, events, or instructions) for further analysis.
Data Transmission:
The monitor sends the captured transactions to:
The Scoreboard to compare actual DUT output with the expected results.
The Coverage Collector to measure functional coverage and identify untested scenarios~\cite{Chakraborty:2002}.
Non-Intrusive Behavior:
The monitor is a passive component, meaning it does not interact or interfere with the DUT operation. Its sole job is to observe and capture signal activity.

\subsubsection{\textbf{Scoreboard in UVM}}
In UVM (Universal Verification Methodology), the Scoreboard is a crucial component that verifies the correctness of the DUT by comparing its actual output with the expected output. It acts as a checker that ensures the DUT behaves as intended by comparing transactions or results observed by the Monitor with a golden reference or expected model.

The scoreboard receives the observed transactions (input and output data) through analysis ports from the monitors. It then performs comparisons and reports mismatches, errors, or deviations from the expected behavior.

\textbf{Key Roles of the Scoreboard:} Result Comparison:
The scoreboard compares the DUT's actual outputs with expected results. Expected results can be generated using a reference model or predetermined values.
Error Detection and Reporting:
If mismatches occur, the scoreboard reports errors with detailed information, enabling quick debugging.
Data Validation:
It checks for data integrity and ensures that the DUT's outputs meet the required specifications.
Functional Verification:
The scoreboard confirms that the DUT adheres to the functional requirements by validating different corner cases and scenarios~\cite{TM:2017}.
Performance Measurement:
In advanced cases, the scoreboard can also help measure the DUT's performance metrics, such as throughput or latency.

\textbf{How the Scoreboard Works in UVM Flow:}
The Monitor observes signals and creates transaction-level data.
The monitor sends the observed transactions to the Scoreboard through an analysis port.
The scoreboard retrieves the expected output (from a golden model, a reference queue, or constraints).
The scoreboard compares the actual DUT output with the expected output.
Any mismatches are reported as errors, while correct outputs are logged as matches.

\begin{figure}
    \centering
    \includegraphics[width=1\linewidth]{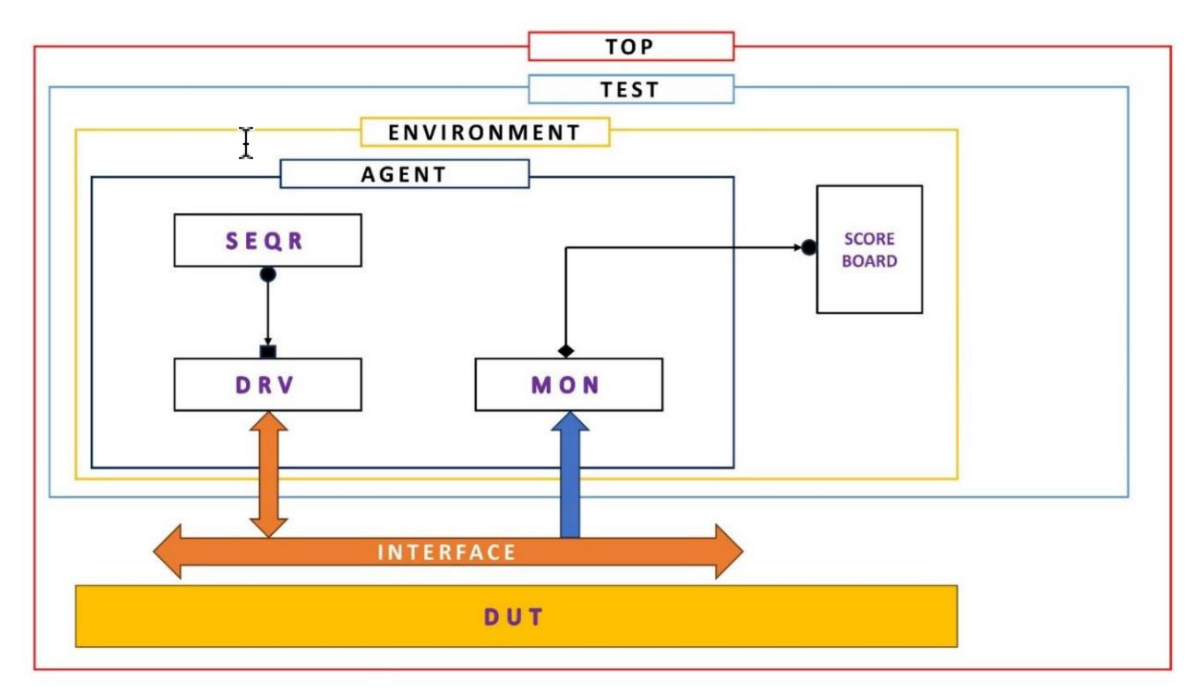}
    \caption{UVM Testbench architecture.}
    \label{fig:enter-label}
\end{figure}

\subsubsection{\textbf{Agent in UVM}}
In UVM (Universal Verification Methodology), the Agent is a verification component that encapsulates and manages the Driver, Sequencer, and Monitor for a specific interface of the Design Under Test (DUT). It acts as a container that bundles these components, providing a complete and modular verification environment for a particular part of the DUT.

The agent simplifies the verification architecture by organizing all the components related to a single interface into a single unit. This modular approach enhances the reusability, configurability, and scalability of the UVM testbench.

\textbf{Structure of an Agent:} A UVM agent typically includes the following components:
Driver: Drives input signals to the DUT by converting transactions into pin-level activity.
Sequencer: Generates and sends transactions (stimulus) to the driver.
Monitor: Passively observes the DUT signals and converts them into transactions for further analysis.
An agent can operate in two modes:
Active Agent: Contains the Driver, Sequencer, and Monitor. It generates and drives stimulus to the DUT.
Passive Agent: Contains only the Monitor. It passively observes DUT activity without driving any signals. It instantiates and connects the Driver, Sequencer, and Monitor during the build phase.

\subsubsection{\textbf{Environment in UVM}}
In UVM (Universal Verification Methodology), the Environment is the top-level container that brings together all the verification components required to verify the DUT. It serves as the central hub of the UVM testbench and is responsible for instantiating, connecting, and configuring components like agents, scoreboards, and coverage collectors.

The environment provides a structured and modular approach to organizing the verification setup, ensuring that components interact seamlessly to verify the functionality of the DUT effectively. It promotes reusability and scalability, allowing the same environment to be adapted for different DUTs or extended for complex verification tasks~\cite{Karthikeyan:2020}.

\textbf{Role of the Environment:}
The environment instantiates all verification components, including agents, scoreboards, and coverage collectors.
It connects components like monitors and scoreboards using analysis ports to ensure smooth data flow.
The environment configures agents and other components using UVM configuration mechanisms like the uvm-config-db.
Functional and code coverage metrics are collected within the environment to ensure verification completeness. 
The environment can be reused across multiple projects or extended to accommodate more complex verification tasks.

\subsubsection{\textbf{Test in UVM}}
In UVM (Universal Verification Methodology), the Test is the top-level component that controls and configures the verification environment. The test is responsible for defining specific test scenarios, configuring the UVM environment, and executing sequences to verify the behavior of the DUT.

A UVM test serves as the entry point for simulation and verification. It instantiates the environment, sets up configuration parameters, and starts sequences that generate stimulus for the DUT. By leveraging UVM's modular and reusable architecture, different tests can be created to verify various functional and corner-case scenarios~\cite{Prasad:2019}.

\textbf{Key Roles of the Test:}
The test instantiates the UVM environment, which contains agents, monitors, scoreboards, and coverage collectors.
The test configures components like agents, drivers, monitors, and scoreboards through the uvm-config-db mechanism.
The test starts the sequence (stimulus generation) using the sequencer. Different tests can execute different sequences to verify various aspects of the DUT.
The test controls the verification flow, including simulation runtime and stopping criteria.

\textbf{Tools used:}
Platform: EDA Playground. 
Simulator: Synopsys VCS (Industry-standard tool for UVM / SystemVerilog simulation).
Waveform Viewer: EPWave for viewing signal transitions and debugging.

\subsection{Traditional Verilog Testbench Method}
In the design and verification process of the FIFO, a standalone Verilog testbench was implemented as a fundamental approach to validate its functionality. This method involves creating a self-contained testbench in Verilog to simulate the behavior of the DUT and verify its functional correctness through directed test scenarios and waveform analysis~\cite{Gupta:2014}.

\textbf{Key Steps in Traditional Verilog Testbench:}
A testbench is written in Verilog, which includes
Clock and Reset Generation. The testbench generates the clock signal and reset signal for the DUT.
Input signals such as write enable, read enable, and data inputs are provided to stimulate the DUT.
The outputs, such as data-out, full, and empty flags, are observed to ensure they match the expected behavior.
Specific test cases (directed scenarios) are written to validate key functional aspects of the FIFO such as verifying that data is written into the FIFO correctly.
The Read Operation ensuring that data can be read correctly in the correct order FIFO.
Boundary Conditions testing the FIFO for edge cases such as empty and full states~\cite{Bergeron:2003}.
Simultaneous Read and Write verifying how the FIFO behaves when read and write operations are triggered simultaneously.

The simulation results are analyzed using waveforms generated by tools like AMD Vivado.
Signals such as clock, reset, data-in, data-out, wn, rn, full, and empty are observed over time to ensure the design behaves as expected.
Waveforms provide a visual representation of how the FIFO responds to stimulus, helping identify any functional issues~\cite{Collett:2018}.

\textbf{Tools Used:} 
AMD Vivado:
AMD Vivado is an advanced FPGA design tool suite used for simulation, synthesis, and implementation of digital designs.
In this project, Vivado was used to simulate the Verilog testbench and generate waveforms to analyze the FIFO's functionality.

\subsection{FPGA Implementation Method}

Similar to conventional approaches~\cite{Croteau:2024, Kiriakidis:2022}, the FPGA implementation is an essential step in verifying a design like a Synchronous FIFO under real-time conditions. While simulation-based verification ensures functional correctness in a software environment, FPGA-based testing evaluates the design in actual hardware, providing insight into timing behavior, practical performance, and real-world functionality.

\textbf{Importantance of FPGA-Based Verification:} The FPGA implementation allows the FIFO design to operate in a real-world environment with actual clocks, resets, and signal inputs, which is not achievable in pure simulation~\cite{IntelFPGA:2020}. Verifying timing parameters like setup time, hold time, and propagation delays is critical. FPGA testing ensures the design performs as expected at the specified clock frequency.
The actual latency, throughput, and response time of the FIFO can be measured, validating the design's performance under realistic conditions.
Besides, FPGA testing enables integration of the FIFO with other system components to evaluate its interaction within a complete design.
FPGA platforms offer on-chip debugging tools, such as the Integrated Logic Analyzer (ILA), to observe internal signals during real-time operation, making it easier to identify issues~\cite{Nagarajan:2018}.

\textbf{Tools and Hardware Used:}
FPGA Board: Intel Altera DE1-SOC Board with Cyclone V FPGA.
Design and Programming Tools:
Intel Quartus Prime: For synthesis, implementation, and timing analysis.
Quartus Programmer: To program the FPGA with the bitstream file.
Debugging Tool:
SignalTap II Logic Analyzer: To observe and debug internal signals in real-time

\begin{figure}
    \centering
    \includegraphics[width=1\linewidth]{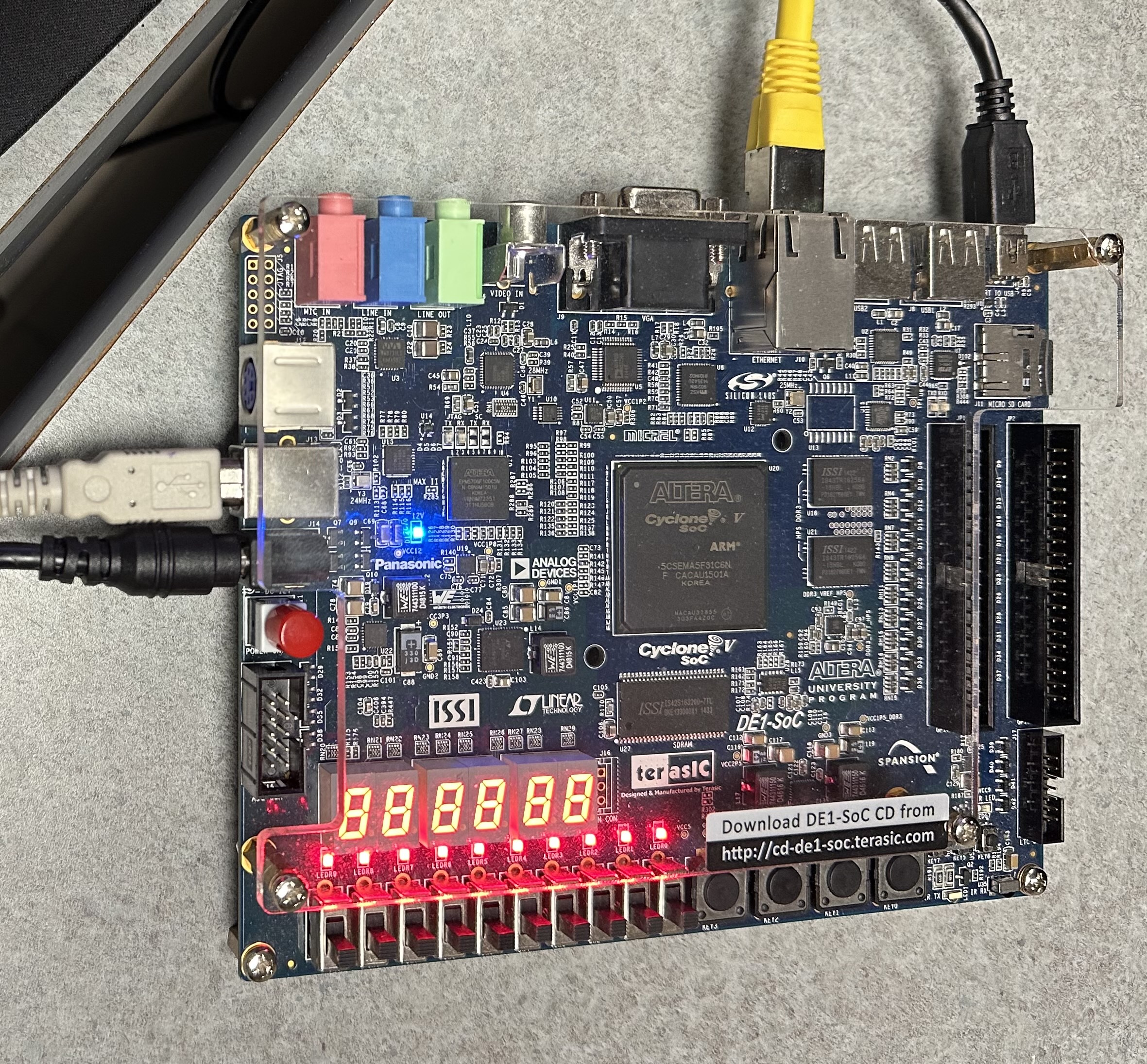}
    \caption{Altera FPGA board used for this project.}
    \label{fig:enter-label}
\end{figure}

\section{Implementation Results and Comparisons}
\subsection{Experimental Setup}
\subsubsection{\textbf{Experimental setup for UVM}}
The Universal Verification Methodology (UVM) testbench was developed, executed, and verified on EDA Playground, an online platform for simulating SystemVerilog/UVM-based designs~\cite{apoorva2020uvm}. The Synopsys VCS simulator was used for running the simulation, and EPWave was utilized for waveform visualization to analyze the behavior of the Synchronous FIFO.

The Implementation steps involved are: 
1. Tool Setup: Platform: EDA Playground 
Simulator: Synopsys VCS (Industry-standard tool for UVM/SystemVerilog simulation).
Waveform Viewer: EPWave (for viewing signal transitions and debugging).

2. The DUT: The Synchronous FIFO RTL design was written in Verilog and instantiated in the UVM testbench. The design supports basic FIFO operations such as read, write operations ~\cite{Tahsildar:2023}

3. UVM Testbench Components: The UVM environment was set up using the following components:
Interface: A SystemVerilog interface was created to connect the DUT with the UVM testbench. It encapsulated signals like clock, reset, data-in, data-out, wn, rn, full, and empty.~\cite{Anusha:2021}
Package: All UVM components (sequence, sequencer, driver, monitor, scoreboard, environment, test) were included in a package.sv file.
UVM Components in order: 

Sequence and Sequencer: 
Generated stimulus (input transactions) for the DUT.

Driver: Converted high-level transactions into pin-level stimulus and drove it to the DUT via the interface.

Monitor: Passively observed DUT signals and converted them into transactions.

Scoreboard: Compared the actual outputs from the DUT with expected outputs for correctness.

Environment: Connected all UVM components.

Test: Top-level testbench that instantiated the environment and triggered the sequence.

Testbench Top:
The tb-top.sv file instantiated the DUT and interface and connected it to the UVM environment. The test was executed using the run-test() method~\cite{Chaudhary:2018}.

Dumpfile Commands:
EPWave requires proper VCD dumping to visualize waveforms

\textbf{Execution Steps:}

Code Compilation:
The design and testbench were compiled using the Synopsys VCS simulator available on EDA Playground.

Simulation:
The run-test method was executed to start the UVM testbench.
UVM messages, including debug information and transaction details, were logged to the simulation output.
Waveform Generation:

The dump.vcd file was generated by including the dumpfile and dumpvars commands. The EPWave viewer on EDA Playground was used to view and analyze the waveform.

\subsubsection{\textbf{Experimental setup for Traditional Verilog Testbench method}}
In this project, AMD Vivado was utilized to simulate the Verilog testbench and produce waveforms for analyzing the functionality of the FIFO~\cite{Patel:2020}.

{\bf Steps Performed in Vivado:} Design Import: The Verilog RTL code for the FIFO and the standalone testbench were imported into Vivado.
Simulation Setup: Simulation scripts and clock/reset configurations were defined.
Execution: Directed test scenarios were run, and the tool simulated the FIFO design.
Waveform Generation: The simulation produced waveforms, enabling verification engineers to visually analyze the FIFO's behavior.
Result Analysis: The outputs, such as data-out, full, and empty, were checked for correctness under various test scenarios. 

\subsubsection{\textbf{Experimental Setup for FPGA Implementation}}

The tools and hardware used for the project include the Intel Altera DE1-SOC Board featuring the Cyclone V FPGA for implementing the design.~\cite{T_Cao_2010} Intel Quartus Prime was employed for synthesis, implementation, and timing analysis of the design, while the Quartus Programmer was used to program the FPGA with the generated bitstream file. For real-time debugging and signal observation, the SignalTap II Logic Analyzer was utilized, enabling the capture and analysis of internal signals to verify and troubleshoot the design's functionality effectively.

\textbf{Steps Involved in FPGA Implementation:}
Design Synthesis: The Verilog RTL code for the FIFO was synthesized using Intel Quartus Prime, the FPGA design tool suite for Intel FPGAs.
Synthesis converts the Verilog description into logic gates and connections that can be mapped onto the programmable hardware resources of the DE1-SOC board.
Design Placement and Routing: The synthesized netlist was placed and routed onto the Altera Cyclone V FPGA on the DE1-SOC board.
The Quartus tool optimized the design for area and timing constraints, ensuring the design met the required clock frequency.
Bitstream Generation: The synthesized design was compiled into a bitstream file that configures the hardware logic on the FPGA.
Programming the FPGA: The bitstream was uploaded to the Intel Altera DE1-SOC FPGA board using the Quartus Programmer tool.
The board was configured to operate the FIFO design under real-time testing conditions.
Real-Time Testing: The design was stimulated by providing input signals such as data-in, wn (write enable), rn (read enable), reset, and clock.
The output signals, such as data-out, full, and empty flags, were observed to ensure the design's correct behavior.
Timing Behavior Analysis: Static Timing Analysis (STA) was performed using Quartus Prime to ensure that timing constraints such as setup time and hold time were met.
The FIFO was tested at various clock frequencies to evaluate its real-time performance.
Signal Debugging and Verification: SignalTap II Logic Analyzer, an on-chip debugging tool available in Quartus Prime, was used to capture and observe internal signals like read/write pointers, data flow, and control flags.
This allowed for real-time signal monitoring and debugging during hardware operation.

\textbf{Example Testing Scenario:}
Write Operation: Data was written into the FIFO using the wn signal and ensure the FIFO correctly entered the data.
Read Operation: Data was read using the rn signal, and ensure that the FIFO correctly read the data.~\cite{Croteau:2024}
Simultaneous Read/Write: The FIFO was tested with simultaneous read and write operations to check for correct behavior without data loss.

\subsection{Results}
\textbf{UVM Result:} From the Figure 4, the output waveform provides insights into the behavior and operation of the synchronous FIFO design. Below is an explanation of the key signals and their interactions:

Clock: The clock signal is a periodic waveform that drives the synchronous FIFO. All operations such as read and write occur on the rising edge of the clock.
Reset: The reset signal is active high during the initial period (around 0 to 100 ns).
When reset is high, all pointers (wptr and rptr), flags (empty and full), and the data outputs (data-out) are initialized to zero.
Write Operation (wn and wptr): The wn signal (write enable) controls when data is written into the FIFO. When wn is high, data is written into the memory location pointed to by the wptr (write pointer).
From the waveform: The wptr increments with every rising edge of the clock when wn is active.
The data-in values (e.g., 26, 19, 17, 08, etc.) are stored into the FIFO memory as indicated by the changes in the wptr~\cite{Gupta:2021}.
Read Operation (rn and rptr): The rn signal controls when data is read from the FIFO. When rn is high, data is read from the memory location pointed to by the rptr.
From the waveform: The rptr increments when rn is active.
The values written into the FIFO (data-in) are sequentially output to data-out, as shown when rn is active.
Empty Flag: The empty flag indicates when the FIFO is empty.
From the waveform: Initially, the empty flag is high during reset, indicating the FIFO is empty.
As data is written (with wn active), the empty flag goes low.
When all written data is read (with rn active), the empty flag goes high again, indicating the FIFO is empty.
Full Flag: The full flag signals when the FIFO is completely filled, and no more writes can occur until some data is read~\cite{apoorva2020uvm}.
From the waveform:
As data is continuously written without being read, the full flag goes high, stopping further writes.
Once data is read (with rn active), the full flag goes low, allowing writes to resume.

Data Flow (data-in and data-out): data-in represents the input data being written into the FIFO when wn is active.
data-out is the output data read from the FIFO when rn is active.
From the waveform: Input data values like 26, 19, 17, etc. are written into the FIFO sequentially.
Correspondingly, the same values are output on data-out during read operations, verifying correct data storage and retrieval.
Therefore, This verifies that the synchronous FIFO is functioning as expected.

\begin{figure}
    \centering
    \includegraphics[width=1\linewidth]{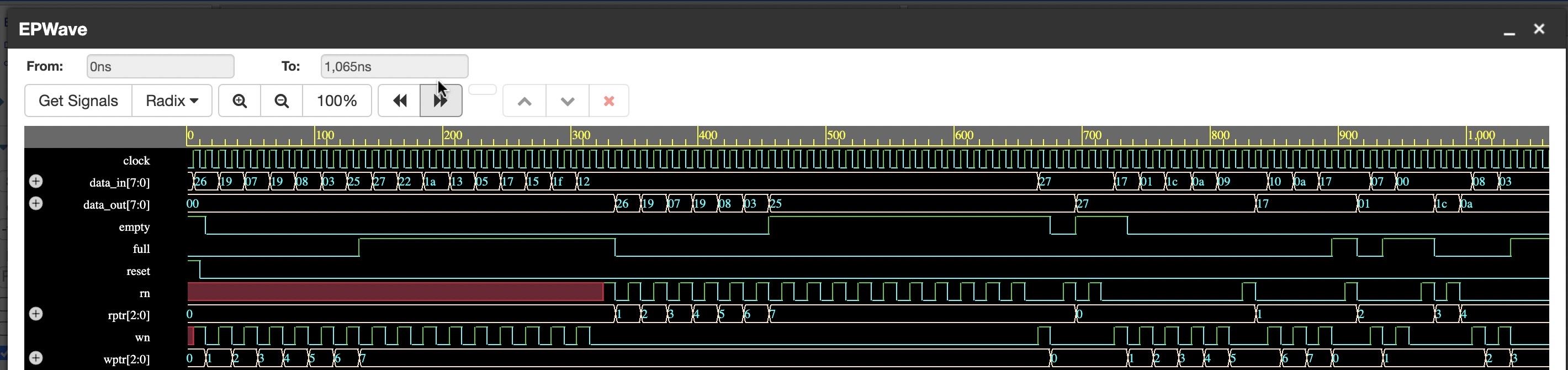}
    \caption{EP wave of Synchronus FIFO in UVM.}
    \label{fig:enter-label}
\end{figure}

\textbf{Traditional Verilog Testbench method Result:}
This is the output waveform of Synchronus FIFO we got in AMD Vivado. 
Similarly, like the previous waveform, this waveform also demonstrates the operation of a synchronous FIFO by showing the behavior of key signals such as clock, reset, wn (write enable), rn (read enable), data-in, data-out, and the status flags full and empty.

Clock and Reset: The clock signal provides the timing reference for the entire FIFO operation, where both write and read operations occur on the rising edge of the clock.
The reset signal initializes the FIFO, setting the data-out, empty flag, and the internal pointers (write pointer wptr and read pointer rptr) to a known state.
Write Operation (wn, data-in): When write enable (wn) is asserted, and the FIFO is not full (full = 0), data is written into the FIFO memory.
From the waveform: Data-in values like 00, a1, b2, c3, d4, e5, f6, 07 are written into the FIFO sequentially as the wn signal goes high.
The write pointer increments with each write operation, storing data in consecutive memory locations.
Read Operation (rn, data-out): When read enable (rn) is asserted, and the FIFO is not empty (empty = 0), the data is read from the FIFO memory.
From the waveform: The data-out signal outputs the data previously written, starting from 00, a1, b2, c3, etc. in sequence.
The read pointer (rptr) increments with each read operation, ensuring sequential data retrieval.
Empty and Full Flags: The empty flag (empty) indicates that the FIFO has no valid data to read. Initially, the FIFO starts as empty (empty = 1).
As data is written into the FIFO, the empty flag goes low, indicating that data is available for reading.
The full flag (full) is asserted when the FIFO becomes completely filled. This prevents further write operations until some data is read.
From the waveform: The empty flag is high at the start and goes low as data is written.
The full flag is observed towards the end of the sequence, indicating that all memory locations have been filled.
The waveform confirms correct write-read behavior, pointer management, and flag operations in a synchronized manner with the clock, verifying its expected operation.

\begin{figure}
    \centering
    \includegraphics[width=1\linewidth]{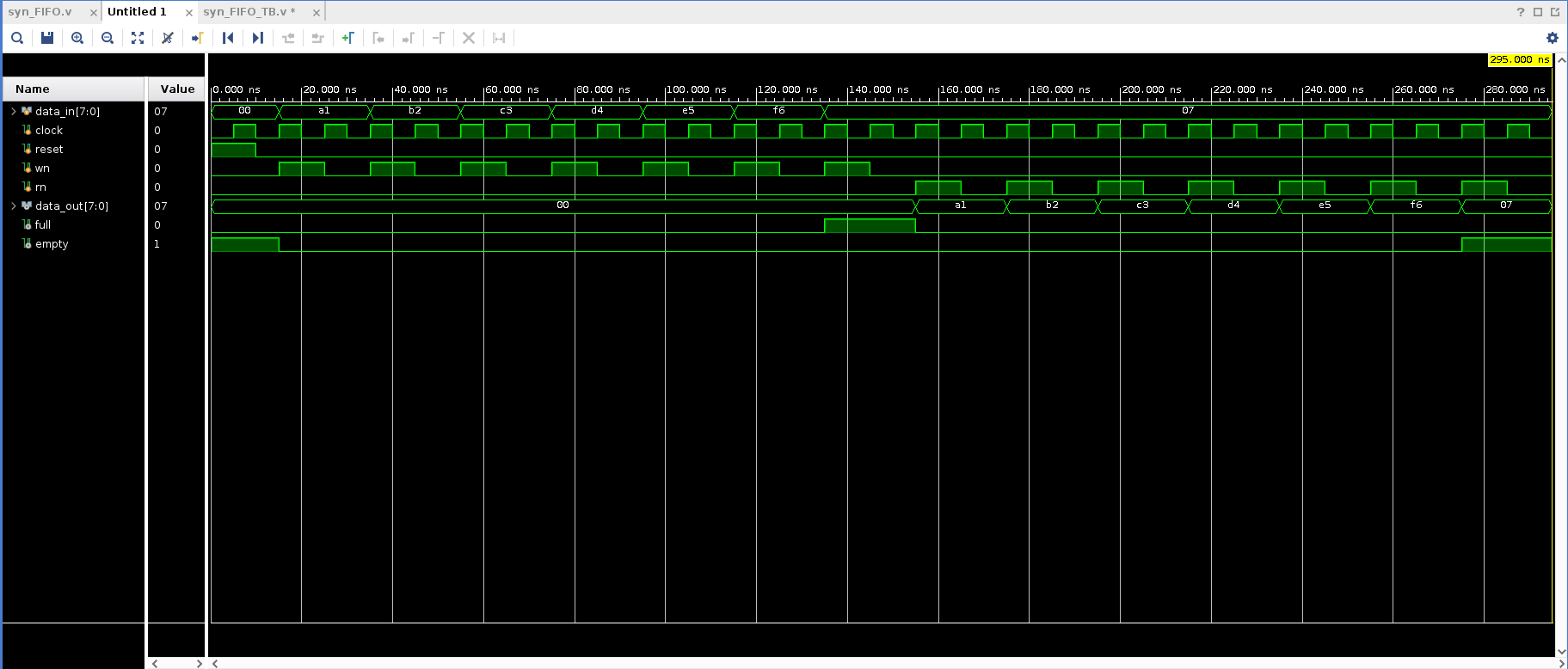}
    \caption{Behavioural Simulation of Synchronus FIFO in Vivado.}
    \label{fig:enter-label}
\end{figure}

\textbf{FPGA Results:} The write operation was performed by asserting the wn (write enable) signal, which allowed data to be written into the FIFO. The design was verified to ensure that the FIFO correctly stored the input data at the appropriate location without any errors~\cite{T_Cao_2010}.
The Command which is used here to write the data is 'mw ff200020 7' where mw is command to write operation and ff200020 is the location where the data will be stored.
The read operation was executed by asserting the rn signal, enabling data to be read from the FIFO. 
The functionality was validated to confirm that the FIFO output the correct data. The Command which is used here to read the data is 'md ff200020 7' where mw is command to read operation and ff200020 is the location from  where the data will be read.

\begin{figure}
    \centering
    \includegraphics[width=1\linewidth]{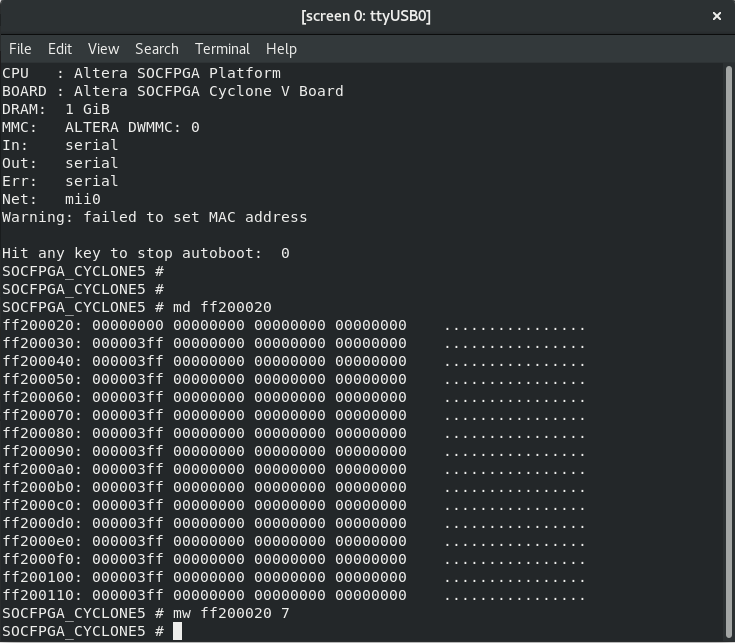}
    \caption{Data storing during write operation.}
    \label{fig:enter-label}
\end{figure}

\begin{figure}
    \centering
    \includegraphics[width=1\linewidth]{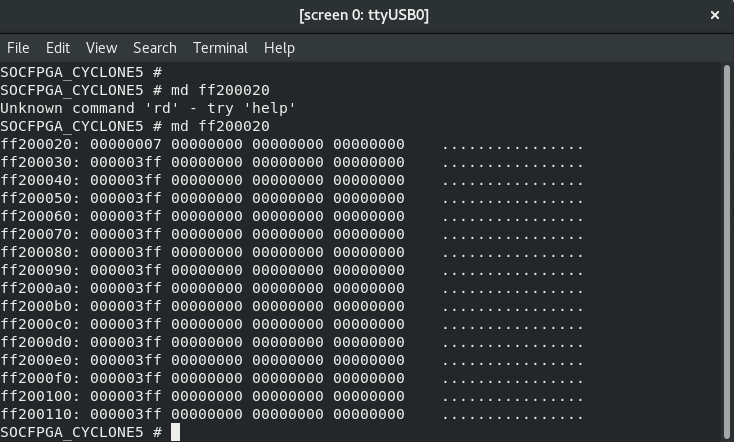}
    \caption{Data reading during read operation.}
    \label{fig:enter-label}
\end{figure}

\textbf{Power Analysis Results:}

The following summarizes the power details of the design:

Cell Internal Power: 10.3991 uW 

Net Switching Power: 34.1859 uW 

Total Dynamic Power: 44.5850 uW

Cell Leakage Power: 86.7468 uW

\textbf{Area report for Synchronus FIFO:}

The following summarizes the area details of the design:

Combinational Area (um2) 1337.051582

Buf/Inv Area (um2) 52.099520

Noncombinational Area (um2) 2092.113444

Net Interconnect Area (um2) 1097.116662

Total Cell Area (um2) 3429.165027

Total Area (um2) 4526.281689

\section{Conclusion}
  
In this project, a synchronous FIFO was designed, verified, and implemented successfully to ensure its functionality and timing behavior. The FIFO design was implemented using Verilog RTL to meet the functional and timing requirements. For verification, multiple approaches were used, including a traditional Verilog testbench and UVM-based verification methodology. These methods helped validate the correctness of the FIFO design under various test scenarios and conditions.  

Additionally, the design was implemented on an Intel Altera DE1-SOC FPGA board with Cyclone V for real-time hardware testing. The FPGA-based implementation verified the performance of the FIFO under practical conditions, ensuring it met the design specifications. Tools such as Vivado were used to simulate the Verilog testbench, analyze waveforms, and confirm the sequential read and write operations, as well as the behavior of the full and empty flags.  

The synchronous FIFO plays a critical role in VLSI systems by efficiently handling data transfer between components operating on the same clock. The successful completion of this project highlights the importance of FIFO in real-time applications, such as data buffering, flow control, and clock domain management.  

Overall, the design, simulation, and hardware implementation demonstrate that the synchronous FIFO operates correctly and reliably, meeting all functional and performance requirements. This project also highlights the value of UVM-based verification and FPGA prototyping for ensuring robust and error-free digital designs.


\ifCLASSOPTIONcompsoc
  \section*{}
\else
  \section*{}
\fi



%

\bibliographystyle{elsarticle-num} 
\bibliography{main.bib}

\begin{thebibliography}{10}
\expandafter\ifx\csname url\endcsname\relax
  \def\url#1{\texttt{#1}}\fi
\expandafter\ifx\csname urlprefix\endcsname\relax\def\urlprefix{URL }\fi
\expandafter\ifx\csname href\endcsname\relax
  \def\href#1#2{#2} \def\path#1{#1}\fi

\bibitem{xie_enze:23}
E.~Xie, J.~Zhou, Analysis and comparison of asynchronous fifo and synchronous fifo, in: 2023 IEEE 2nd International Conference on Electrical Engineering, Big Data and Algorithms (EEBDA), 2023, pp. 260--264.
\newblock \href {https://doi.org/10.1109/EEBDA56825.2023.10090586} {\path{doi:10.1109/EEBDA56825.2023.10090586}}.

\bibitem{Ghosh:2023}
A.~Ghosh, D.~Mitra, A comparative study of asynchronous and synchronous fifos for embedded systems, Microelectronics Journal 134 (2023) 105872.
\newblock \href {https://doi.org/10.1016/j.mejo.2023.105872} {\path{doi:10.1016/j.mejo.2023.105872}}.

\bibitem{Ciletti:2010}
M.~Ciletti, Advanced Digital Design with Verilog HDL, 2nd Edition, Prentice Hall, 2010.

\bibitem{Islam_diff:2015}
R.~Islam, H.~Fahmy, P.-Y. Lin, M.~R. Guthaus, {Differential current-mode clock distribution}, in: 2015 IEEE 58th International Midwest Symposium on Circuits and Systems (MWSCAS), 2015, pp. 1--4.
\newblock \href {https://doi.org/10.1109/MWSCAS.2015.7282042} {\path{doi:10.1109/MWSCAS.2015.7282042}}.

\bibitem{Kumar:2014}
A.~Kumar, N.~Sharma, Verification of asynchronous fifo using system verilog, International Journal of Computer Applications 86~(11) (2014) 1--5.
\newblock \href {https://doi.org/10.5120/15095-3665} {\path{doi:10.5120/15095-3665}}.

\bibitem{islam2018low}
R.~Islam, {Low-power resonant clocking using soft error robust energy recovery flip-flops}, Journal of Electronic Testing 34~(4) (2018) 471--485.

\bibitem{guthaus2017current}
M.~Guthaus, R.~Islam, {Current-mode clock distribution}, uS Patent 9,787,293 (Oct.~10 2017).

\bibitem{Islam_neg:2021}
R.~Islam, {Negative Capacitance Clock Distribution}, IEEE Transactions on Emerging Topics in Computing 9~(1) (2021) 547--553.
\newblock \href {https://doi.org/10.1109/TETC.2018.2872000} {\path{doi:10.1109/TETC.2018.2872000}}.

\bibitem{Islam_hcdn:2019}
R.~Islam, M.~R. Guthaus, {HCDN: Hybrid-Mode Clock Distribution Networks}, IEEE Transactions on Circuits and Systems I: Regular Papers 66~(1) (2019) 251--262.
\newblock \href {https://doi.org/10.1109/TCSI.2018.2866224} {\path{doi:10.1109/TCSI.2018.2866224}}.

\bibitem{Islam_dcmcs:2018}
R.~Islam, H.~A. Fahmy, P.~Y. Lin, M.~R. Guthaus, {DCMCS: Highly Robust Low-Power Differential Current-Mode Clocking and Synthesis}, IEEE Transactions on Very Large Scale Integration (VLSI) Systems 26~(10) (2018) 2108--2117.

\bibitem{Lin:2015}
P.-Y. Lin, H.~A. Fahmy, R.~Islam, M.~R. Guthaus, {LC resonant clock resource minimization using compensation capacitance}, in: 2015 IEEE International Symposium on Circuits and Systems (ISCAS), 2015, pp. 1406--1409.

\bibitem{islam2017current}
R.~Islam, {Current-Mode Clocking and Synthesis Considering Low-Power and Skew}, University of California, Santa Cruz, 2017.

\bibitem{Croteau:2024}
B.~Croteau, K.~Kiriakidis, T.~A. Severson, R.~Robucci, S.~Rahman, R.~Islam, {State Estimation Adaptable to Cyberattack Using a Hardware Programmable Bank of Kalman Filters}, IEEE Transactions on Control Systems Technology 32~(5) (2024) 1730--1742.
\newblock \href {https://doi.org/10.1109/TCST.2024.3378991} {\path{doi:10.1109/TCST.2024.3378991}}.

\bibitem{saini2024reconfigurable}
R.~Saini, R.~Islam, {Reconfigurable can intrusion detection and response system}, Electronics 13~(13) (2024) 2672.

\bibitem{Kiriakidis:2022}
K.~Kiriakidis, B.~Croteau, T.~Severson, E.~Rodriguez-Seda, R.~Robucci, R.~Islam, S.~Rahman, Degradable tracking system based on hardware multi-model estimators, in: 2022 Resilience Week (RWS), 2022, pp. 1--6.
\newblock \href {https://doi.org/10.1109/RWS55399.2022.9984042} {\path{doi:10.1109/RWS55399.2022.9984042}}.

\bibitem{islam2022systems}
R.~Islam, {Systems and methods of graph-based vehicular intrusion detection}, uS Patent App. 17/713,369 (Nov.~24 2022).

\bibitem{Narkhede:2019}
N.~Narkhede, A.~Mohini, {Design and Verification of Generic FIFO using Layered Test Bench and Assertion Technique}, International Journal of Engineering and Advanced Technology (IJEAT) 8~(6) (2019) 5254--5260.

\bibitem{Nanda:2018}
M.~Nanda, {Implementation and Verification of Asynchronous FIFO Under Boundary Condition}, International Journal of Engineering Research and Technology 7~(4) (2018) 1--6.

\bibitem{Islam_sedu:2019}
R.~Islam, {Low-Power Highly Reliable SET-Induced Dual-Node Upset-Hardened Latch and Flip-Flop}, Canadian Journal of Electrical and Computer Engineering 42~(2) (2019) 93--101.
\newblock \href {https://doi.org/10.1109/CJECE.2019.2895047} {\path{doi:10.1109/CJECE.2019.2895047}}.

\bibitem{islam2011high}
R.~Islam, {High-speed energy-efficient soft error tolerant flip-flops}, Ph.D. thesis, Concordia University (2011).

\bibitem{Islam_sdff:2011}
R.~Islam, S.~Esmaeili, T.~Islam, {A high performance clock precharge SEU hardened flip-flop}, in: 2011 9th IEEE International Conference on ASIC, 2011, pp. 574--577.
\newblock \href {https://doi.org/10.1109/ASICON.2011.6157270} {\path{doi:10.1109/ASICON.2011.6157270}}.

\bibitem{Esmaeili:2012}
S.~E. Esmaeili, R.~Islam, A.~J. Al-Khalili, G.~E.~R. Cowan, {Dual-edge triggered sense amplifier flip-flop utilizing an improved scheme to reduce area, power, and complexity}, in: 2012 19th IEEE International Conference on Electronics, Circuits, and Systems (ICECS 2012), 2012, pp. 292--295.
\newblock \href {https://doi.org/10.1109/ICECS.2012.6463565} {\path{doi:10.1109/ICECS.2012.6463565}}.

\bibitem{Islam_isqed:2012}
R.~Islam, {A highly reliable SEU hardened latch and high performance SEU hardened flip-flop}, in: Thirteenth International Symposium on Quality Electronic Design (ISQED), 2012, pp. 347--352.
\newblock \href {https://doi.org/10.1109/ISQED.2012.6187516} {\path{doi:10.1109/ISQED.2012.6187516}}.

\bibitem{Bhanu:2021}
C.~A. Bhanu, R.~Odela, G.~R. Padmini, Coverage driven verification of synchronous fifo using uvm, {Recent Trends in Electronics and Communication Systems} 8~(1) (2021) 1--7.

\bibitem{5993629}
D.~Li, P.~I.-J. Chuang, D.~Nairn, M.~Sachdev, {Design and analysis of metastable-hardened flip-flops in sub-threshold region}, in: IEEE/ACM International Symposium on Low Power Electronics and Design, 2011, pp. 157--162.
\newblock \href {https://doi.org/10.1109/ISLPED.2011.5993629} {\path{doi:10.1109/ISLPED.2011.5993629}}.

\bibitem{Rizvi:2015}
N.~Z. Rizvi, R.~Arora, \href{https://www.ojs.jctecs.com/index.php/com/article/download/19/12}{Implementation and verification of synchronous fifo using system verilog verification methodology}, Journal of Communications Technology, Electronics and Computer Science 2 (2015) 18--23.
\newline\urlprefix\url{https://www.ojs.jctecs.com/index.php/com/article/download/19/12}

\bibitem{Sharma:2013}
H.~Sharma, C.~Rana, {Designing of 8-bit Synchronous FIFO Memory using Register File}, International Journal of Scientific and Research Publications 3~(5) (2013) 1--5.

\bibitem{islam2022feasibility}
R.~Islam, {Feasibility prediction for rapid IC design space exploration}, Electronics 11~(7) (2022) 1161.

\bibitem{Islam_early:2022}
R.~Islam, {Early Stage DRC Prediction Using Ensemble Machine Learning Algorithms}, IEEE Canadian Journal of Electrical and Computer Engineering 45~(4) (2022) 354--364.
\newblock \href {https://doi.org/10.1109/ICJECE.2022.3200075} {\path{doi:10.1109/ICJECE.2022.3200075}}.

\bibitem{challagundla2023design}
D.~Challagundla, I.~Bezzam, R.~Islam, {Design automation of series resonance clocking in 14-nm FinFETs}, Circuits, Systems, and Signal Processing 42~(12) (2023) 7549--7579.

\bibitem{islam2019predicting}
R.~Islam, M.~A. Shahjalal, {Predicting DRC violations using ensemble random forest algorithm}, in: Proceedings of the 56th Annual Design Automation Conference 2019, 2019, pp. 1--2.

\bibitem{Wyland:1993}
D.~Wyland, New features in synchronous fifos, in: Proceedings of WESCON '93, 1993, pp. 580--585.
\newblock \href {https://doi.org/10.1109/WESCON.1993.488598} {\path{doi:10.1109/WESCON.1993.488598}}.

\bibitem{islam2021high}
R.~Islam, {High-speed on-chip signaling: Voltage or current-mode?}, IETE Journal of Research 67~(2) (2021) 217--226.

\bibitem{Maurya:2016}
S.~Maurya, \href{https://www.ijert.org/design-of-rtl-synthesizable-32-bit-fifo-memory}{Design of rtl synthesizable 32-bit fifo memory}, International Journal of Engineering Research \& Technology (IJERT) 5~(11) (2016) 591--593.
\newline\urlprefix\url{https://www.ijert.org/design-of-rtl-synthesizable-32-bit-fifo-memory}

\bibitem{Darda:2015}
A.~Darda, {Implementation and Verification of Synchronous FIFO using System Verilog Verification Methodology}, International Journal of Scientific \& Engineering Research 6~(7) (2015) 123--128.

\bibitem{Ahmed:2023}
S.~E. Ahmed, \href{https://github.com/SeiffAhmed22/Synchronous-FIFO-UVM-Verification}{Synchronous fifo uvm verification}, GitHub Repository (2023).
\newline\urlprefix\url{https://github.com/SeiffAhmed22/Synchronous-FIFO-UVM-Verification}

\bibitem{Schirrmeister:2014}
F.~Schirrmeister, \href{https://ieeexplore.ieee.org/document/6800481}{Panel: Future soc verification methodology: Uvm evolution or revolution?}, Design, Automation \& Test in Europe Conference \& Exhibition (DATE) (2014) 1--6.
\newline\urlprefix\url{https://ieeexplore.ieee.org/document/6800481}

\bibitem{Darbari:2016}
A.~Darbari, I.~Singleton, \href{https://arxiv.org/abs/1606.02347}{Industrial strength formal using abstractions}, arXiv preprint arXiv:1606.02347 (2016).
\newline\urlprefix\url{https://arxiv.org/abs/1606.02347}

\bibitem{Jain:2022}
R.~Jain, K.~Sharma, High-speed synchronous fifo design with integrated clock gating techniques, International Journal of VLSI Design and Communication Systems 13~(2) (2022) 123--134.
\newblock \href {https://doi.org/10.5121/vlsidesign.2022.13209} {\path{doi:10.5121/vlsidesign.2022.13209}}.

\bibitem{Liu:2021}
Y.~Liu, Q.~Wang, H.~Zhang, Design and optimization of synchronous fifos in multi-clock domains for high-speed communication, IEEE Access 9 (2021) 112345--112356.
\newblock \href {https://doi.org/10.1109/ACCESS.2021.3105678} {\path{doi:10.1109/ACCESS.2021.3105678}}.

\bibitem{Saxena:2018}
A.~Saxena, D.~Gupta, Functional verification of fifo design using systemverilog and coverage analysis, in: 2018 International Conference on Advances in Computing, Communication Control and Networking (ICACCCN), 2018, pp. 203--208.
\newblock \href {https://doi.org/10.1109/ICACCCN.2018.8748432} {\path{doi:10.1109/ICACCCN.2018.8748432}}.

\bibitem{Cheng:2019}
L.~Cheng, W.~Zhu, Efficient design of synchronous fifo with low latency and high throughput, Journal of Circuits, Systems, and Computers 28~(3) (2019) 1950043.
\newblock \href {https://doi.org/10.1142/S0218126619500431} {\path{doi:10.1142/S0218126619500431}}.

\bibitem{Juan:2014}
J.~Francesconi, J.~Agustin~Rodriguez, P.~M. Julián, Uvm based testbench architecture for unit verification, in: 2014 Argentine Conference on Micro-Nanoelectronics, Technology and Applications (EAMTA), 2014, pp. 89--94.
\newblock \href {https://doi.org/10.1109/EAMTA.2014.6906085} {\path{doi:10.1109/EAMTA.2014.6906085}}.

\bibitem{IEEE1800:2017}
I.~S. Association, Ieee 1800.2-2017, IEEE Standard for Universal Verification Methodology (UVM) (2017) 1--220\href {https://doi.org/10.1109/IEEESTD.2017.7921364} {\path{doi:10.1109/IEEESTD.2017.7921364}}.

\bibitem{Mehta:2013}
A.~B. Mehta, System verilog assertions and functional coverage, Springer (2013).
\newblock \href {https://doi.org/10.1007/978-1-4614-6491-1} {\path{doi:10.1007/978-1-4614-6491-1}}.

\bibitem{Kadam:2017}
A.~V. Kadam, P.~S. Kadam, Functional verification of fifo using systemverilog and uvm, International Journal of Engineering Development and Research (2017) 215--220.

\bibitem{Agarwal:2014}
N.~Agarwal, Implementation and verification of fifo using system verilog methodology, International Journal of Computer Applications 86~(11) (2014) 1--5.
\newblock \href {https://doi.org/10.5120/15095-3665} {\path{doi:10.5120/15095-3665}}.

\bibitem{Chakraborty:2002}
S.~Chakraborty, \href{https://www.date-conference.com/proceedings-archive/2017/pyear/PAPERS/2002/DATE02/PDFFILES/TUTORIAL1.PDF}{Tutorial one functional verification of system on chips - practices, issues and challenges}, Design, Automation \& Test in Europe Conference \& Exhibition (DATE) (2002) 1--6.
\newline\urlprefix\url{https://www.date-conference.com/proceedings-archive/2017/pyear/PAPERS/2002/DATE02/PDFFILES/TUTORIAL1.PDF}

\bibitem{TM:2017}
T.~M. Pavithran, R.~Bhakthavatchalu, Uvm based testbench architecture for logic sub-system verification, in: 2017 International Conference on Technological Advancements in Power and Energy ( TAP Energy), 2017, pp. 1--5.
\newblock \href {https://doi.org/10.1109/TAPENERGY.2017.8397323} {\path{doi:10.1109/TAPENERGY.2017.8397323}}.

\bibitem{Karthikeyan:2020}
S.~Karthikeyan, A.~Venkatraman, Uvm verification of a configurable fifo for real-time applications, International Journal of Electronics 107~(4) (2020) 561--574.
\newblock \href {https://doi.org/10.1080/00207217.2020.1712193} {\path{doi:10.1080/00207217.2020.1712193}}.

\bibitem{Prasad:2019}
R.~Prasad, P.~Sharma, Synchronous fifo design and verification for data stream buffers, in: 2019 IEEE International Workshop on Integrated Circuits (IWIC), 2019, pp. 120--124.
\newblock \href {https://doi.org/10.1109/IWIC.2019.9025667} {\path{doi:10.1109/IWIC.2019.9025667}}.

\bibitem{Gupta:2014}
R.~A. Gupta, Design and implementation of synchronous fifo using verilog hdl, International Journal of Advanced Research in Computer and Communication Engineering 3~(9) (2014) 67--71.

\bibitem{Bergeron:2003}
J.~Bergeron, Writing Testbenches: Functional Verification of HDL Models, Springer, 2003.

\bibitem{Collett:2018}
M.~Collett, D.~Richmond, Using uvm for advanced verification of synchronous fifos, IEEE Design Automation Conference (DAC) (2018) 135--140.

\bibitem{IntelFPGA:2020}
I.~FPGA, \href{https://www.intel.com/}{Design and verification of fifo with altera fpgas}, Intel FPGA Resource Center (2020).
\newline\urlprefix\url{https://www.intel.com/}

\bibitem{Nagarajan:2018}
V.~Nagarajan, \href{https://repository.rit.edu/theses/9980/}{The design and verification of a synchronous first-in first-out (fifo) module using system verilog based universal verification methodology (uvm)}, International Journal of Scientific Research in Science, Engineering and Technology 5~(1) (2018) 199--204.
\newline\urlprefix\url{https://repository.rit.edu/theses/9980/}

\bibitem{apoorva2020uvm}
A.~H. M, K.~Bailey, Uvm based design verification of fifo, International Journal of Engineering Research \& Technology (IJERT) 9~(6) (2020) 1--5.

\bibitem{Tahsildar:2023}
D.~Tahsildar, \href{https://github.com/Daniyal-Tahsildar/FIFO_TB_UVM}{Fifo tb uvm: Implementation of a synchronous fifo with uvm testbench}, GitHub Repository (2023).
\newline\urlprefix\url{https://github.com/Daniyal-Tahsildar/FIFO_TB_UVM}

\bibitem{Anusha:2021}
C.~A. Bhanu, R.~Odela, G.~R. Padmini, \href{https://engineeringjournals.stmjournals.in/index.php/RTECS/article/view/5494}{Coverage driven verification of synchronous fifo using uvm}, Recent Trends in Electronics and Communication Systems 8~(1) (2021) 1--7.
\newline\urlprefix\url{https://engineeringjournals.stmjournals.in/index.php/RTECS/article/view/5494}

\bibitem{Chaudhary:2018}
A.~Chaudhary, R.~Singh, Design and implementation of high-performance synchronous fifos for fpga-based systems, ACM Transactions on Design Automation of Electronic Systems 23~(4) (2018) 40.
\newblock \href {https://doi.org/10.1145/3200657} {\path{doi:10.1145/3200657}}.

\bibitem{Patel:2020}
R.~Patel, P.~Kumar, Uvm-based verification of low-power fifo for iot applications, in: 2020 IEEE International Symposium on Circuits and Systems (ISCAS), 2020, pp. 1--4.
\newblock \href {https://doi.org/10.1109/ISCAS.2020.9151876} {\path{doi:10.1109/ISCAS.2020.9151876}}.

\bibitem{T_Cao_2010}
T.~Cao, J.~Chang, D.~Gong, C.~Liu, T.~Liu, A.~Xiang, J.~Ye, \href{https://dx.doi.org/10.1088/1748-0221/5/12/C12003}{Design and verification of a bit error rate tester in altera fpga}, Journal of Instrumentation 5~(12) (2010) C12003.
\newblock \href {https://doi.org/10.1088/1748-0221/5/12/C12003} {\path{doi:10.1088/1748-0221/5/12/C12003}}.
\newline\urlprefix\url{https://dx.doi.org/10.1088/1748-0221/5/12/C12003}

\bibitem{Gupta:2021}
R.~Gupta, A.~Kumar, Implementation of area-efficient fifo design for resource-constrained environments, IEEE Embedded Systems Letters 13~(1) (2021) 14--17.
\newblock \href {https://doi.org/10.1109/LES.2021.3054569} {\path{doi:10.1109/LES.2021.3054569}}.

\end{thebibliography}


\begin{thebibliography}{1}
\bibitem{IEEEhowto:kopka}
V. Sharma and S. Saini, "Design and Implementation of Synchronous FIFO Using Verilog HDL for High-Speed Data Processing", IEEE International Conference on Computing for Sustainable Global Development (INDIACom), 2019, pp. 789-794.

\bibitem{IEEEhowto:kopka}
J. Park and K. S. Kim, "Performance Analysis and Optimization of Synchronous FIFO Design in ASICs and FPGAs", IEEE Transactions on Very Large Scale Integration (VLSI) Systems, vol. 25, no. 4, 2017, pp. 1212-1220.

\bibitem{IEEEhowto:kopka}
A. S. Kumar and N. B. Aravind, "Optimized Design and Verification of Synchronous FIFO Using UVM Methodology", IEEE International Symposium on VLSI Design and Test (VDAT), 2018, pp. 99-104.

\bibitem{IEEEhowto:kopka}
C. Lee and W. Chiu, "Design and Verification of Multi-Depth Synchronous FIFO for Data Buffers", IEEE International Workshop on Electronics, Control, Measurement, Signals, and Their Applications (WECMSA), 2015, pp. 251-256.

\bibitem{IEEEhowto:kopka}
R. Mehta, “Efficient Design and Verification of FIFO Buffers,” Journal of VLSI Design and Verification, Vol. 12, No. 4, 2018.

\bibitem{IEEEhowto:kopka}
J. Bergeron, E. Cerny, A. Hunter, and A. Nightingale, “Verification
Methodology Manual for System Verilog.”, Springer, 2006.

\bibitem{IEEEhowto:kopka}
Wang, Xin, Tapani Ahonen, and Jari Nurmi, "A synthesizable RTL
design of asynchronous FIFO”, System-on-Chip Proceedings. 2004
International Symposium on. IEEE, pp. 123-128, 2004.

\bibitem{IEEEhowto:kopka}
S. Vijayaraghavan and M. Ramanathan, “A Practical Guide for System
Verilog Assertions”. Springer, 2005

\bibitem{Croteau:2024}
Croteau, Brien and Kiriakidis, Kiriakos and Severson, Tracie A. and Robucci, Ryan and Rahman, Saad and Islam, Riadul, “State Estimation Adaptable to Cyberattack Using a Hardware Programmable Bank of Kalman Filters”. IEEE Transactions on Control Systems Technology, Vol. 32, No. 5, 2024.

@ARTICLE{10490234,
  author={Croteau, Brien and Kiriakidis, Kiriakos and Severson, Tracie A. and Robucci, Ryan and Rahman, Saad and Islam, Riadul},
  journal={IEEE Transactions on Control Systems Technology}, 
  title={State Estimation Adaptable to Cyberattack Using a Hardware Programmable Bank of Kalman Filters}, 
  year={2024},
  volume={32},
  number={5},
  pages={1730-1742},
  keywords={Field programmable gate arrays;Hardware;Cyberattack;Program processors;State estimation;Radar tracking;Fabrics;Cyber-physical systems;Decentralized control;Resilience;Cyber-physical security;distributed control;resilient state estimation;robust adaptive control},
  doi={10.1109/TCST.2024.3378991}}




\end{thebibliography}

\end{document}